\documentclass{article}

    \PassOptionsToPackage{numbers}{natbib}

\usepackage[final]{neurips_2022}

\usepackage[utf8]{inputenc} 
\usepackage[T1]{fontenc}    
\usepackage{hyperref}       
\usepackage{url}            
\usepackage{booktabs}       
\usepackage{amsfonts}       
\usepackage{nicefrac}       
\usepackage{microtype}      
\usepackage{xcolor}         
\usepackage{bm}
\usepackage{amsmath,amsfonts,amssymb}
\usepackage{threeparttable}
\usepackage{subcaption}
\usepackage{graphicx}
\title{Exploring General Intelligence via Gated Graph Transformer in Functional Connectivity Studies}

\author{%
  Gang~Qu\\
  Department of Biomedical Engineering\\
  Tulane University\\
  \texttt{gqu1@tulane.edu} \\
  \And
  Anton Orlichenko \\
  Tulane University \\
  \texttt{aorlichenko@tulane.edu} \\
  \AND
  Junqi Wang \\
  Department of Radiology\\
  Cincinnati Children’s Hospital Medical Center \\
  \texttt{junqi.wang@cchmc.org} \\
  \And
  Gemeng Zhang \\
  Department of Neurology\\
  Mayo Clinic\\
  \texttt{zhang.gemeng@mayo.edu} \\
  \And
  Li Xiao \\
  School of Information Science and Technology\\
  University of Science and Technology of China and Institute of Artificial Intelligence \\
  \texttt{xiaoli11@ustc.edu.cn} \\
  \And
  Aiying Zhang\\
  School of Data Science\\
  University of Virginia\\
  \texttt{aiying.zhang@virginia.edu} \\
  \And
  Zhengming Ding \\
  Department of Computer Science\\
  Tulane University \\
  \texttt{zding1@tulane.edu} \\
  \And
  Yu-Ping Wang\\
  Department of Biomedical Engineering\\
  Tulane University\\
  \texttt{wyp@tulane.edu} \\
}

\begin{document}

\maketitle

\begin{abstract}
Functional connectivity (FC) as derived from fMRI has emerged as a pivotal tool in elucidating the intricacies of various psychiatric disorders and delineating the neural pathways that underpin cognitive and behavioral dynamics inherent to the human brain. While Graph Neural Networks (GNNs) offer a structured approach to represent neuroimaging data, they are limited by their need for a predefined graph structure to depict associations between brain regions, a detail not solely provided by FCs. To bridge this gap, we introduce the Gated Graph Transformer (GGT) framework, designed to predict cognitive metrics based on FCs. Empirical validation on the Philadelphia Neurodevelopmental Cohort (PNC) underscores the superior predictive prowess of our model, further accentuating its potential in identifying pivotal neural connectivities that correlate with human cognitive processes.
\end{abstract}
\section{Introduction}
Functional magnetic resonance imaging (fMRI) is a powerful, non-invasive tool that detects changes in the brain's blood-oxygenation-level-dependent (BOLD) signal, offering insights into the functional organization of the brain. This technique has proven instrumental in studying neurological and psychiatric disorders like autism, Alzheimer's disease, and schizophrenia\citep{cherkassky2006functional, sheline2013resting, QUMGCN2021, Anton10002422}. The concept of functional connectivity (FC) has emerged as a potential fingerprint for cognitive and behavioral traits, spurring interest in modeling and extracting these connectivity patterns to propel neuroscience research forward.

Graph Neural Networks (GNNs)\citep{kipf2017semi} are at the forefront of analyzing brain FC, effectively distilling meaningful information from complex neuroimaging connectivity patterns. They have been pivotal in tasks like forecasting cognitive abilities and detecting brain disorders\citep{Yansurvey, QuGCN2021}. However, the challenge lies in pinpointing the appropriate graph structure that illustrates connections between brain regions. While transformer GNNs\citep{dwivedi2021generalization, kreuzer2021rethinking, ying2021transformers} have been suggested to overcome this, they face intrinsic limitations, including model interpretability issues and reduced performance in scenarios with weak connectivity between brain regions.

Addressing these challenges, we introduce a novel framework based on the Gated Graph Transformer (GGT) to predict individual cognitive abilities using FCs. This model employs a refined random-walk diffusion strategy, learnable structural and positional encodings, and a gating mechanism to enhance the learning process\citep{dwivedi2021graph}. Crucially, the self-attention mechanism in our framework discerns intricate relationships between brain regions while maintaining computational efficiency. We validate our model on the Philadelphia Neurodevelopmental Cohort (PNC)\citep{satterthwaite2014neuroimaging} to predict the Wide Range Achievement Test (WRAT) score, a measure of general intelligence\citep{kareken1995reading}, with results surpassing various baseline models. This research underscores the importance of blending spatial brain network knowledge with advanced neural network techniques, laying the groundwork for future neuroscience research.

\section{Methods}
We propose an end-to-end GGT model, combining a learnable position encoding\citep{dwivedi2021graph, 10363771} with brain network structure to boost the predictive performance. Let $\mathcal{G}=(\mathcal{V},\mathcal{E})$ be a graph, in which nodes and edges represent the brain regions of interest (ROIs) and the associated connections between ROIs, respectively. We assume $\bm{E}^l\in\mathbb{R}^{n_\mathcal{E}\times d_\mathcal{E}}$ to be the edge features, $\bm{H}^l\in\mathbb{R}^{n_\mathcal{V}\times d_\mathcal{V}}$ to be the node features, and $\bm{P}^l\in\mathbb{R}^{n_\mathcal{V}\times d_p}$ to be the node position embeddings on at the $l_{th}$ layer of GGT, where $n_\mathcal{V}$ and $n_\mathcal{E}$ are the number of nodes and edges on the graph, and $d_\mathcal{V}$, $d_\mathcal{E}$, and $d_p$ are the feature dimensions. We first learn 3 embedding weights $\bm{Q}^l, \bm{K}^l, \bm{V}^l\in \mathbb{R}^{d^{l}\times d^{l+1}}$, where $d^{l}$ and $d^{l+1}$ are the input and output feature dimensions, respectively. GGT propagates the connection information using the attention mechanism, as defined in Eq.\ref{attention}, to focus on important neighbors, enabling the model to learn complex relationships between nodes.
\begin{equation}
\alpha_{ij}^l=\frac{\mathrm{exp}(\mathrm{tanh}(\bm{Q}^l\begin{bmatrix}\bm{H}_i^l\\\bm{P}_i^l\end{bmatrix})\cdot \mathrm{sig}(\bm{K}^l\begin{bmatrix}\bm{H}_j^l\\\bm{P}_j^l\end{bmatrix})}{\sum_{i\in N_j}\mathrm{exp}(\mathrm{tanh}(\bm{Q}^l\begin{bmatrix}\bm{H}_i^l\\\bm{P}_i^l\end{bmatrix})\cdot \mathrm{sig}(\bm{K}^l\begin{bmatrix}\bm{H}_k^l\\\bm{P}_k^l\end{bmatrix}))},\label{attention}
\end{equation}
where $\mathrm{sig}(\cdot)$ and $\mathrm{tanh}(\cdot)$ are the sigmoid and hyperbolic tangent activation functions, respectively. Specifically, the attention score is computed to measure the similarity of the query($\bm{Q}$) and key($\bm{K}$) mappings, indicating how much attention is given to a particular neighbor of each node. The GGT, as formulated in Eq.\ref{propagation}, utilizes a framework analogous to the gated control system. This framework combines edge features with position encoding, operates directly on the edges, and thereby modulates the importance of the messages being propagated across the graph..
\begin{align}
    &\bm{H}^{l+1}_i= \bm{H}^{l}_i+\mathrm{ReLU}(\bm{V}^l\begin{bmatrix}\bm{H}_k^l\\\bm{P}_k^l\end{bmatrix}+\sum_{j\in N_i} \eta_{ij}^l\odot (\alpha_{ij}^l\bm{V}^l\begin{bmatrix}\bm{H}_j^l\\\bm{P}_j^l\end{bmatrix})),\nonumber\\
    &\bm{P}_i^{l+1}=\bm{P}_i^l+\mathrm{tanh}(\bm{C}_1^l \bm{P}_i^l+\sum_{j\in N_i} \eta_{ij}^l\odot \bm{C}_2^l\bm{P}_j^l), \qquad \bm{E}_{ij}^{l+1}=\bm{E}_{ij}^l+\mathrm{ReLU}(\hat{\eta}^l_{ij}),
    \label{propagation}
\end{align}
 where $\eta_{ij}^l=\frac{\mathrm{sig}(\hat{\eta}_{ij}^l)}{\sum_{j^\prime\in N_j}\mathrm{sig}(\hat{\eta}_{ij^\prime}^l)+\epsilon}$ and $\hat{\eta}_{ij}^l=\bm{B}_1^l\bm{H}_i^l+\bm{B}_2^l\bm{H}_j^l+\bm{B}_3^l\bm{E}_{ij}^l$; $\bm{C}_1^l, \bm{C}_2^l, \bm{B}_1^l, \bm{B}_2^l\in \mathbb{R}^{d_\mathcal{V}\times d_\mathcal{V}}$, $\bm{B}_3^l\in\mathbb{R}^{d_\mathcal{E}\times d_\mathcal{E}}$ are the learnable weights. The position encoding $\bm{P}^0$ is initialized based on the random walk (RW) diffusion process, which incorporates both the spatial information of MNI (Montreal Neurological Institute) coordinates and frequency information of the FC graph. 
 
$\bm{P}^0=[RW, RW^2, \cdots,RW^{d_p-3}, MNI_x, MNI_y, MNI_z]$, where $RW=\bm{A}\bm{D}^{-1}$, and $MNI_x$, $MNI_y$, $MNI_z$  represent the MNI coordinates. The matrices $\bm{A}\in\mathbb{R}^{n_\mathcal{V}\times n_\mathcal{V}}$ and $\bm{D}\in\mathbb{R}^{n_\mathcal{V}\times n_\mathcal{V}}$ are the adjacency matrix and the degree matrix, respectively.
Next, a multilayer perceptron (MLP) uses the mean graph pooling as the input for the label prediction. The loss function can be written as $\mathrm{Loss}=\mathrm{MSE}(\hat{y}, y)+\lambda_1 \mathrm{trace}(\bm{P}^\mathrm{T}L\bm{P})+\lambda_2 \lVert \bm{P}\bm{P}^\mathrm{T}-I\rVert_F^2$, where $\hat{y}$ and $y$ are the prediction and label, $\bm{L}=\bm{D}^{-\frac{1}{2}}\bm{A}\bm{D}^{-\frac{1}{2}}$ is the graph Laplacian, and $\lambda_{1,2}>0$ are hyperparameters controlling the relative importance of position encoding, which balance smoothness and orthonormality of the learned PE vectors at the final GGT layer.

\section{Experiments}
We evaluate our method using two paradigms of fMRI (emotion identification (emoid) task and working memory task (nback)) from PNC.
All BOLD scans are acquired on a single $3$T Siemens TIM Trio whole-body scanner with a single-shot, interleaved multi-slice, gradient-echo, echo-planar imaging sequence. The total scanning time is $50$ minutes, $32$ seconds (TR $=3000$ ms, TE $=32$ ms, and flip angle $=90$ degrees), with a voxel resolution of $3$ mm and $46$ slices. Gradient performance is $45$ mT/m, with a maximum slew rate of $200$ T/m/s.
\begin{table*}[htb!]
  \begin{center}
    \caption{The performance of prediction on WRAT scores with different models.}
    \label{regres}
    \begin{threeparttable}
    \begin{tabular}{l|c|c|c|c}
    \hline
      \textbf{Model} &\textbf{Emoid RMSE(mean $\pm$ std) }&\textbf{Nback RMSE(mean $\pm$ std) }\\
      \hline
LR      &  22.9844 $\pm$ 1.4313    & 22.9632 $\pm$ 1.1070    \\
\hline
GCN     &  15.6924 $\pm$ 0.9912    & 15.4335 $\pm$ 0.5867   \\
\hline
MTL     &  16.1742 $\pm$ 0.9283    & 16.2187 $\pm$ 0.9576   \\
\hline
M2TL    &  15.9967 $\pm$ 0.8947  &15.0807 $\pm$ 0.8902   \\
\hline
NM2TL   &  15.0320 $\pm$ 0.7965   & 15.0310 $\pm$ 0.7955  \\
\hline
\textbf{GGT}     &  \textbf{14.5530 $\pm$ 0.7454}  &  \textbf{14.7961 $\pm$ 0.5284}   \\
\hline
    \end{tabular}
    \end{threeparttable}
  \end{center}
\end{table*}
SPM12\footnote{\url{http://www.fil.ion.ucl.ac.uk/spm/software/spm12/}} is used to conduct motion correction, spatial normalization, and smoothing with a 3mm Gaussian kernel. Subject fMRI scans are parcellated according to the Power264 atlas\citep{power2011functional}. We use the FCs derived from Pearson's correlation of the fMRI as the initial node features and combine the sparse FC matrix (i.e., only keep the $K=30$ largest neighbors of each node) with the MNI distance matrix as the edge features.

We randomly split the data into training, validation, and test sets with the ratio of $70\%$, $10\%$, and $20\%$, respectively. The model is trained on the training set with the hyperparameters tuned on the validation set using random search. We compare our model with linear regression (LR), Multi-task learning model (MTL)\citep{zhang2012multi}, Manifold regularized multi-task learning (M2TL)\citep{jie2015manifold}, and New Manifold regularized multi-task learning (NM2TL)\citep{xiaoNMTL2022}. The predictive performance is evaluated using bootstrap analysis on $10$ repeated experiments, as shown in Table.\ref{regres}. The results show that the proposed GGT model achieves better predictive performance than other methods on both fMRI paradigms.
\begin{figure}[h!]
  \centering
    \includegraphics[width=1.0\textwidth]{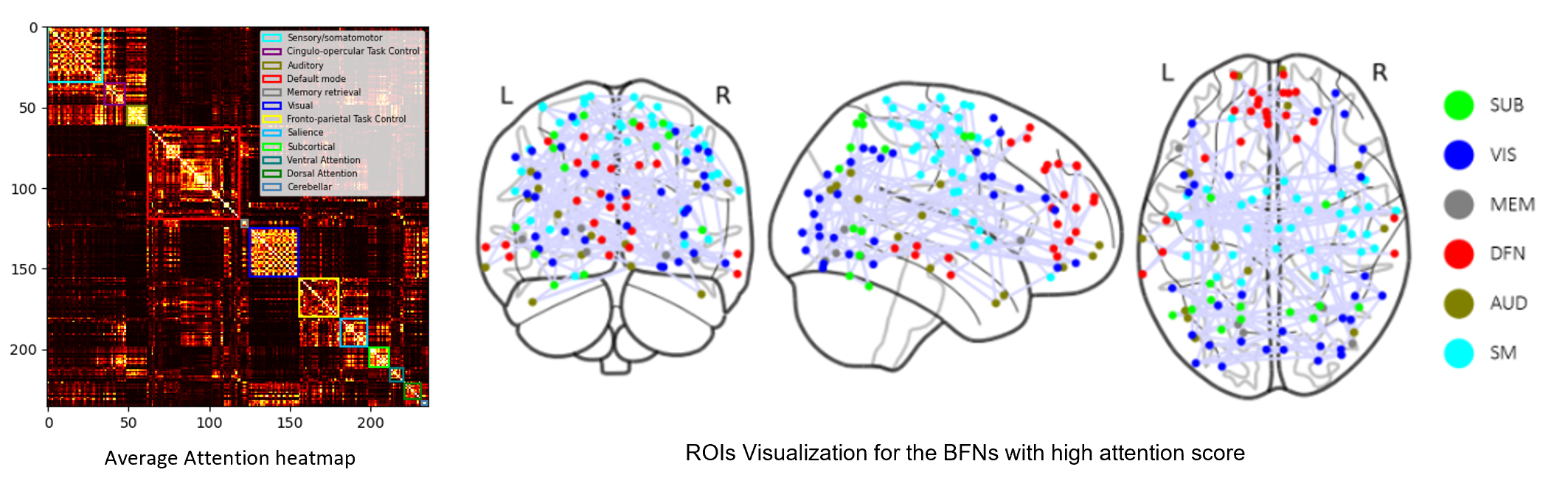}
  \caption{Average heatmap generated using the attention scores (left) and brain functional networks with high attention
scores (right). SUB: Subcortical, VIS: Visual, MEM:
Memory Retrieval, DFN: Default Mode, AUD: Auditory, SM:
Sensory/somatomotor.}
\label{fig:heatmap}
\end{figure}

Additionally, our model can be interpreted through the heatmap produced by the attention scores. It's worth noting that the transposed attention score matrix is added to achieve symmetric results. As shown in Fig.\ref{fig:heatmap}, the WRAT score related brain regions can be identified for two paradigms.
Without prior clustering knowledge, several brain functional networks (BFN)\citep{eickhoff2018imaging} can be identified from the heatmap, including the auditory network (AUD)\citep{westerhausen2010identification}, visual network (VIS)\citep{satterthwaite2016philadelphia}, and memory retrieval network (MEM)\citep{unsworth2014working} in a purely data-driven way. In addition, subsets of FCs in the somatomotor network (SM), default mode network (DFN), and subcortical (SUB) have higher attention scores than other FCs. Moreover, the results are consistent for both paradigms, demonstrating the reliability of our method in identifying significant ROIs.

\section{Conclusion}
\label{sec:conclusion}
In this paper, we propose a novel end-to-end interpretable GGT framework to conduct brain functional connectivity studies. GGT explores both functional and spatial relations between individual brain regions and combines the learnable position encoding in brain network structure to boost predictive performance. We further investigate the model explainability with the attention mechanism by identifying the significant biomarkers underlying general intelligence. The proposed model shows great potential for understanding and predicting the variance in human cognitive function.

\section{Potential Negative Societal Impact}
The authors do not identify any potential negative societal impact for this work on the society. 

\section{Acknowledgements}
This work was supported in part by NIH under Grants R01 GM109068, R01 MH104680, R01 MH107354, P20 GM103472, R01 REB020407, R01 EB006841, and 2U54MD007595, in part by NSF under Grant $\#$1539067, and in part by National Natural Science Foundation of China under Grant No. 62202442 and Anhui Provincial Natural Science Foundation under Grant No. 2208085QF188

\bibliographystyle{unsrtnat}  
\bibliography{main}   

\end{document}